\documentclass[twoside,agupp]{aguplus}  

\usepackage{mathptmx}	      
\usepackage{graphicx}	      
\usepackage[latin1]{inputenc} 
\usepackage[T1]{fontenc}      
\usepackage{xspace}	      

\newcommand{\eg}{\emph{e.g.}}                   
\newcommand{\etc}{\emph{etc}}                   
\newcommand{\viz}{\emph{viz}}                   
\newcommand{\Tromso}{Troms\o\xspace}            
\newcommand{\Sub}[1]{\ensuremath_{\mathrm{#1}}} 
\newcommand{\Te}{T\Sub{e}}                      
\newcommand{\Ti}{T\Sub{i}}

\newcommand{\Vi}{V\Sub{i}}                      
\newcommand{\Ne}{N\Sub{e}}
\newcommand{\Es}{E\Sub{s}}                      
\newcommand{\Bx}{B_x}
\newcommand{\By}{B_y}
\newcommand{\Bz}{B_z}
\newcommand{\coords}[2]{$#1\deg$N, $#2\deg$E}   
\newcommand{\cf}{\emph{cf}.\xspace}             
\newcommand{\hm}{h\Sub{m}}                      
\newcommand{\Xmax}{X\Sub{max}}                  
\newcommand{\Eth}{E\Sub{th}}                    
\renewcommand{\fd}{f\Sub{d}}                    
\newcommand{\SigmaA}{\Sigma\Sub{A}}             
\newcommand{\SigmaP}{\Sigma\Sub{P}}             
\newcommand{\VA}{V\Sub{A}}                      
\newcommand{\Vth}{V\Sub{th}}                    

\printfigures
\doublecaption{35pc}
\extraabstract
\afour
\figmarkoff
\lefthead{Blagoveshchenskaya et al.}
\righthead{HF pump wave triggering of local auroral activation}
\received{}
\revised{}
\accepted{}
\journalid{JGRA}{January 2000}
\articleid{1}{4}
\paperid{99JZ12345}
\ccc{0000-0000/00/99JZ-12345\$05.00}
\cpright{AGU}{1999}
\slugcomment{Submitted to Journal of Geophysical Research, 27 June, 1999.
Resubmitted, 2 May, 2000.}

\title{Ionospheric HF pump wave triggering of local auroral activation}
\author{N.~F.~Blagoveshchenskaya,\altaffilmark{1} V.~A.~Kornienko,\altaffilmark{1}
T.~D.~Borisova,\altaffilmark{1} B.~Thidé,\altaffilmark{2} M.~J.~Kosch,%
\altaffilmark{3} M.~T.~Rietveld,\altaffilmark{3} E. V. Mishin,%
\altaffilmark{4} R.~Yu.~Luk'yanova,\altaffilmark{1} and O.~A.~Troshichev%
\altaffilmark{1}}
\altaffiltext{1}{Arctic and Antarctic Research Institute,
 St.~Petersburg, Russia.
}
\altaffiltext{2}{Swedish Institute of Space Physics,
 Uppsala Division, Sweden.
}
\altaffiltext{3}{Max-Planck-Institut f\"ur Aeronomie,
 Katlenburg-Lindau, Germany.
}
\altaffiltext{4}{MIT Haystack Observatory, Westford,  Mass., USA
}
\authoraddr{
 N.~F.~Blagoveshchenskaya,
 V.~A.~Kornienko,
 T.~D.~Borisova,
 R.~Yu.~Luk'yanova,
 and
 O.~A.~Troshichev,
 Arctic and Antarctic Research Institute,
 38 Bering Street, St.~Petersburg~199\,397, Russia.
 (nataly@aari.nw.ru, vikkorn@aari.nw.ru, olegtro@aari.nw.ru)
}
\authoraddr{B.~Thid\'e,
 Swedish Institute of Space Physics, Uppsala Division,
 SE-755\,91\,Uppsala, Sweden.
 (bt@irfu.se)
}
\authoraddr{
 M.~J.~Kosch
 and
 M.~T.~Rietveld,
 Max-Planck-Institut f\"ur Aeronomie, D-3411 Katlenburg-Lindau, Germany.
(kosch@linax1.mpae.gwdg.de, rietveld@mirage.mpae.gwdg.de)
}
\authoraddr{E.~V.~Mishin,
 MIT Haystack Observatory, Westford, MA01886, USA.
 (evm@haystack.mit.edu)
}

\begin{document}


\begin{abstract}

Experimental results from \Tromso HF pumping experiments in the
nightside auroral $E$ region are reported.  We found intriguing evidence
that a modification of the ionosphere-magnetosphere coupling, due to the
effects of powerful HF waves beamed into an auroral sporadic $E$ layer,
can lead to a local intensification of the auroral activity.
Summarizing multi-instrument ground-based observations and observations
from the IMP\,8 and IMP\,9 satellites, one can distinguish the following
peculiarities related to this auroral activation:  modification of the
auroral arc and its break-up above \Tromso; local changes of the
horizontal currents in the vicinity of \Tromso; increase of the electron
temperature and ion velocities at altitudes above the HF pump reflection
level; distinctive features in dynamic HF radio scatter Doppler spectra;
pump-induced electron precipitation; substorm activation exactly above
\Tromso.  The mechanisms of the modification of the
ionosphere-magnetosphere coupling through the excitation of the
turbulent Alfv\'en boundary layer between the base of the ionosphere
($\sim100$~km) and the level of sharp increase of the Alfv\'en velocity
(at heights up to one Earth radius), and the formation of a local
magnetospheric current system are discussed.  The results suggest that a
possible triggering of local auroral activation requires specific
geophysical conditions.

\end{abstract}


\section{Introduction}
\label{sec:intro}

\begin{sloppypar}
Field-aligned currents (FACs) play an important r\^ole in the process of
energy transfer between the magnetosphere and the ionosphere.  They
establish the force balance between the hot, tenuous plasma of the
magnetosphere and the cold, dense plasma of the ionosphere
\citep{Haerendel90}.  Important agents in the magnetosphere-ionosphere
coupling are Alfv\'en waves, since they are equivalent to time-varying
field-aligned currents.  It has been suggested that the static coupling
is dominant for large-scale magnetosphere-ionosphere coupling, while the
Alfv\'en wave coupling is dominant for small-scale coupling
\citep{Nagatsuma&al96}.  One of the most remarkable manifestations of
the dynamic processes in the solar wind-magnetosphere-iono\-sphere
system is the substorm phenomenon.  It is known that a magnetospheric
substorm involves two components:  the directly driven component, and
the energy storage release part.  The driven system responds directly to
changes in the interplanetary medium (IMF orientation, solar wind
pressure) and involves the direct deposition of solar wind energy into
the auroral ionosphere and the symmetric ring currents.  The
storage-release system pertains to the process of storage of energy in
the tail magnetic field and in the kinetic drift of the magnetotail
particles.  After that, it is explosively released into the auroral
ionosphere and symmetric ring current system
\citep{Rostoker&al87,Rostoker99}.  As pointed out by \citet{Rostoker99}
the progress of the substorm may be produced by either the more global,
directly driven process or a more localized, storage-release process and
may be looked at either on a global scale or on a local scale.
\end{sloppypar}

During a substorm onset, the large-scale laminar magnetospheric
convection is disrupted because the magnetospheric cross-tail current is
diverted down the magnetic field lines \citep{McPherron&al73}. This 
leads to the activation of the substorm current loop, the so-called
substorm current wedge \citep{Kamide&Baumjohann93,Rostoker&al87}.
\citet{Bostrom64} proposed two auroral current systems connected to the
auroral electrojet.  This model has not only remained valid up to
present time, but also comprises most suggestions made later on.

At the ionospheric level the substorm expansion onset is characterized
by brightening and subsequent break-up of a pre-existing auroral arc
\citep{Akasofu64}.  There are indications that some auroral arcs are
generated by field line resonances (FLRs) \citep{Samson&al96}.

\citet{Lui&Murphree98} proposed a substorm onset model by combining a
theory of the auroral arc generation due to FLRs with a theory of
current disruption in the near-Earth magnetic tail based on the
cross-field current instability.  It allows a close tie of current
disruption region in the magnetotail to the location of the auroral arc.

In recent times, it has been generally assumed that the ionosphere plays
a rather passive r\^ole in the substorm process.  Nonetheless, some
exceptions to this assumption exist.  Firstly, models emphasizing
changes in the ionospheric conductivity proposed by \citet{Kan&Sun85},
\citet{Kan93}, and \citet{Lysak90}, and its r\^ole in enhancing the
field-aligned currents \textit{Lysak and Song} [1998].  Secondly, models
which emphasized the decoupling of magnetospheric convection from the
ionosphere, induced by the formation of parallel electric fields
\citep{Haerendel90}.  Lastly, there are models which emphasize the
excitation of the turbulent Alfv\'en boundary layer in the polar
ionosphere, thus giving rise to a strong turbulent heating of the plasma
and to the production of accelerated particles
\citep{Trakhtengerts&Feldstein91}.

To prove the active r\^ole of the auroral ionosphere in the substorm
process, the controlled injection of high-power radio waves into space
from purpose-built ground-based HF radio facilities has constituted an
excellent tool.

Experimental results concerning artificial modification of the
ionosphere-magnetosphere system by HF pump waves were presented by
\citet{Blagoveshchenskaya&al98-AG,Blagoveshchenskaya&al99}.

In this paper we report experimental results from \Tromso HF pumping
experiments in the nightside auroral $E$ region and present evidence for
a modification, produced by powerful HF radio waves, of the
ionosphere-magnetosphere coupling, leading to a local intensification of
the auroral activity.  Data from bistatic HF Doppler radio scatter, the
IMAGE magnetometer network, the EISCAT UHF radar, the \Tromso dynasonde,
the digital all-sky imager (DASI), and the IMP\,8 and IMP\,9 satellites
were used in the analysis.


\section{Experimental methods and equipment used}

The experiments reported here were conducted by using the EISCAT HF
heating facility \citep{Rietveld&al93} located near \Tromso
(geographical coordinates \coords{69.6}{19.2}, $L=6.2$, magnetic dip
angle $I=78\deg$) in the pre-midnight hours of February~16 and 17, 1996.
The \Tromso heater was operating at the frequency $4040$ kHz, $O$-mode
polarization, and an effective radiated power of $150$~MW.  The antenna
beam was tilted $6\deg$ to the south, thus allowing HF pumping in a near
field-aligned direction.

\begin{figure}[t]
 \figurewidth{\linewidth}
 \figbox*{}{}{%
  \includegraphics[width=.95\linewidth]{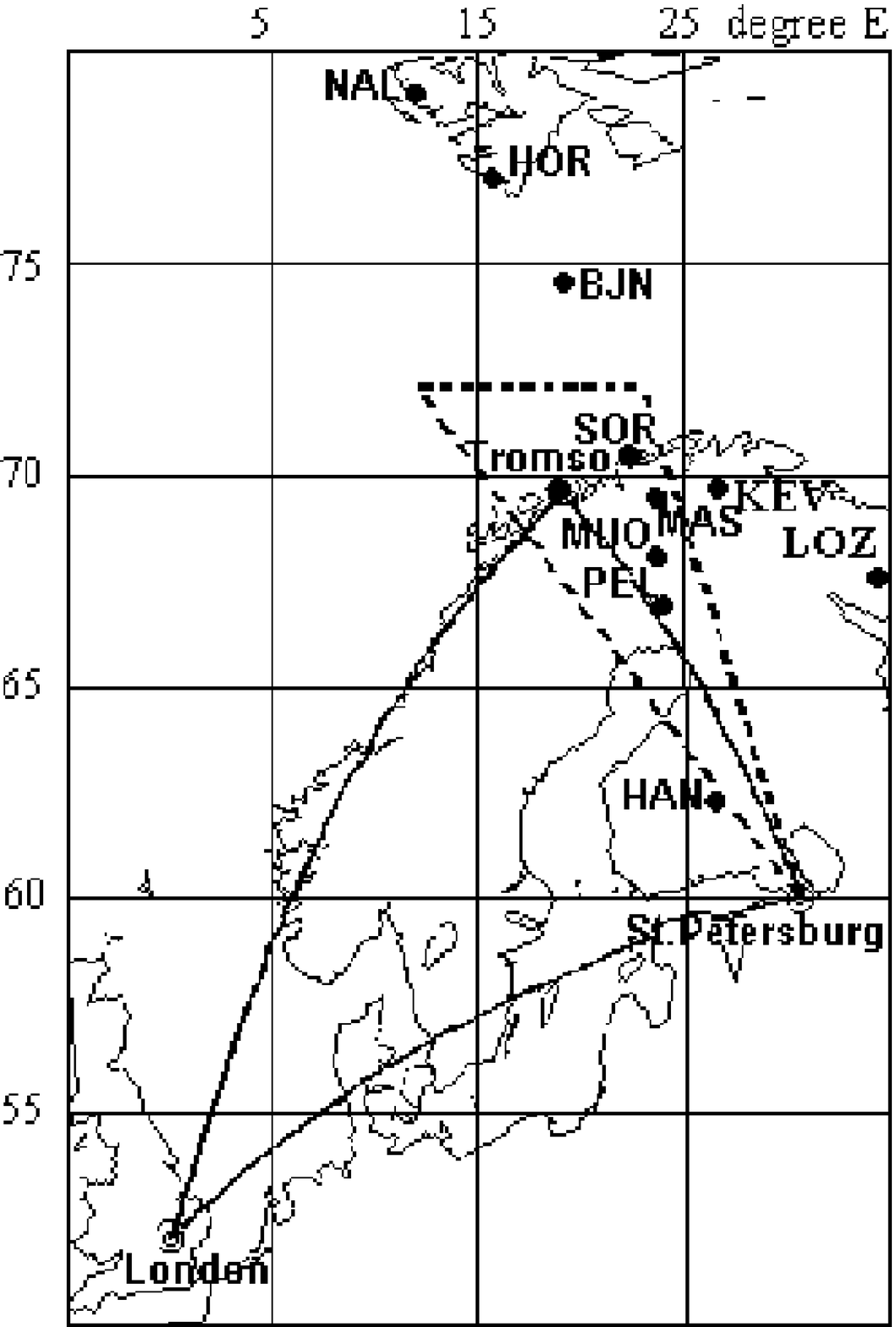}
 }
 \caption{
 General view of the experiment geometry, indicating the position
 of the London--\Tromso--St.~Petersburg path, where HF Doppler measurements of
 the HF signals scattered from AFAIs were made, and the locations of the IMAGE
 magnetometer stations. The names of the stations used are as follows: Ny 
 {\AA}lesund (NAL; \coords{78.92}{11.95}),
 Hornsund (HOR; \coords{77.00}{15.60}),
 Bear Island (BJN; \coords{74.50}{19.20}),
 S{\o}r{\o}ya (SOR; \coords{70.54}{22.22}),
 \Tromso (TRO; \coords{69.66}{18.94}),
 Masi (MAS; \coords{69.46}{23.70}),
 Muonio (MUO; \coords{68.02}{23.53}),
 Pello (PEL; \coords{66.90}{24.08}),
 Hankasalmi (HAN; \coords{62.30}{26.65}),
 Kevo (KEV; \coords{69.76}{17.01}), 
 and
 Lovozero (LOZ; \coords{67.97}{22.22}).
 }
 \label{fig:geometry}
\end{figure}

Bistatic scatter measurements of HF diagnostic signals were carried out
on the London--\Tromso--St.~Petersburg path at operational frequencies
of $9410$ and $12,095$~kHz.  The analysis of the received diagnostic
waves, scattered from artificial field-aligned irregularities (AFAIs)
above \Tromso, was made with a Doppler spectral method in St.~Petersburg
at a distance of about $1200$~km; the receiving antenna was directed
toward \Tromso.  The geometry of the experiments is shown in
Figure~\ref{fig:geometry}.  Spectral processing of the diagnostic
signals was made with a Fast Fourier Transform (FFT) method.  On the 16
and 17 February, the frequency bandwidth used was $50$ and $33$~Hz,
respectively, with a frequency resolution of about $0.1$~Hz and a
temporal resolution of about $10$~s.

To facilitate the interpretation of the Doppler measurements, we used
data from the IMAGE magnetometer network \citep{Luhr&al98}.  The
locations of the IMAGE magnetometers are depicted in
Figure~\ref{fig:geometry}.  The time resolution of the IMAGE
magnetometers used in this study was $10$~s.

The information about disturbances in the interplanetary magnetic field
(IMF) and in the solar wind, that could be related with the substorm
onset, was obtained from IMP\,8 and IMP\,9 satellite data.  The time
resolution of the satellite data used in this study was 1 minute.

Optical data were obtained with the digital all-sky imager (DASI).
Details of the instrumentation can be found in the work by
\citet{Kosch&al98}.  The DASI is located at Skibotn, near \Tromso
(\coords{69.3}{20.4}).  In this study we used $557.7$~nm data with
$10$~s and $30$~s temporal resolution.

To obtain information about changes of the electron densities and
temperatures, $\Ne$ and $\Te$, and ion temperatures and velocities,
$\Ti$ and $\Vi$, during HF pumping experiments, EISCAT UHF radar
measurements at \Tromso were also employed.

\begin{plate*}[p]
 \figurewidth{\linewidth}
 \figbox*{}{}{%
  \includegraphics[clip,viewport=0 350 550 650,width=1.0\linewidth]
  {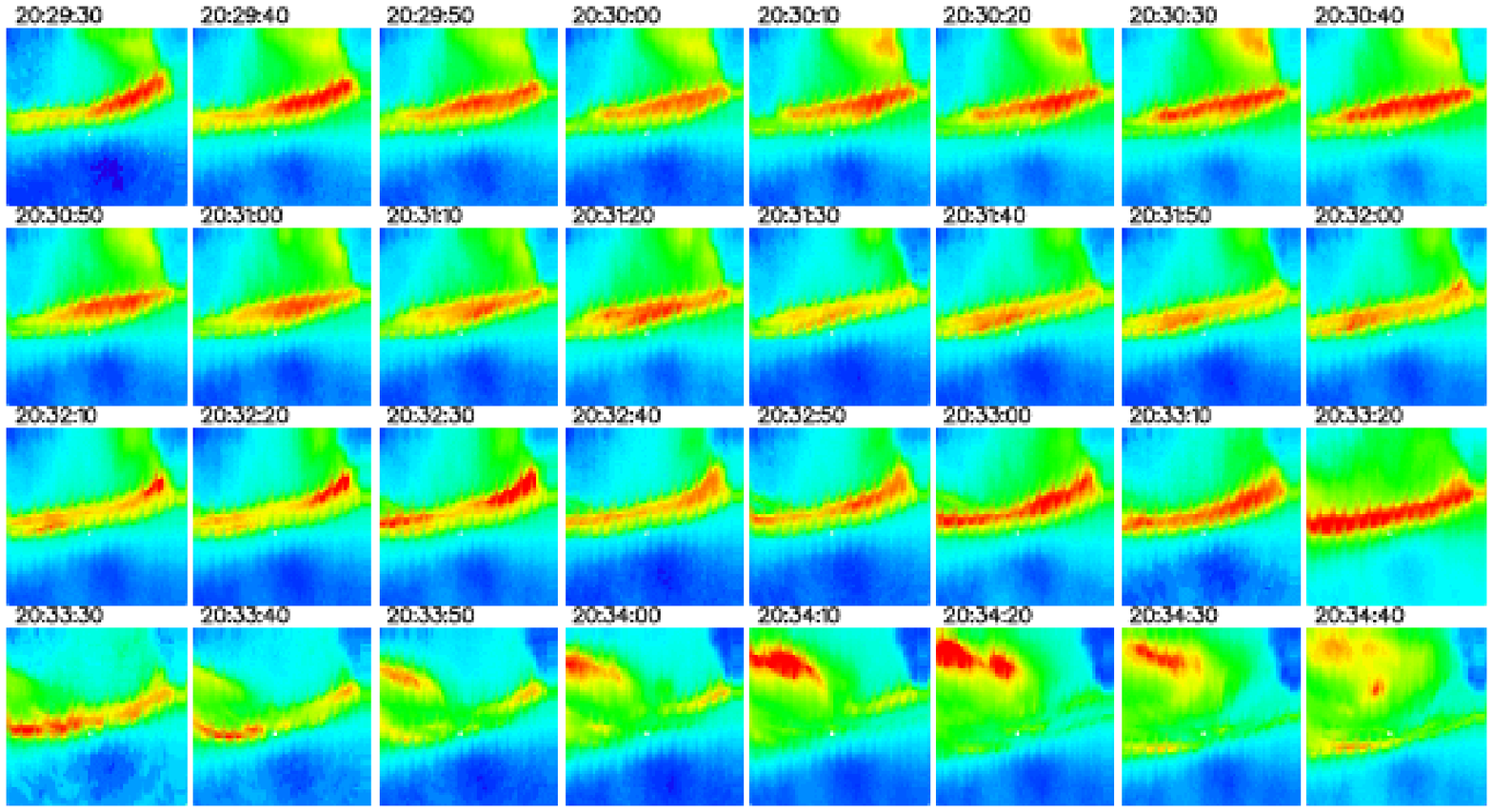}
 }
 \caption{
 A sequence of all-sky imager plots at $557.7$~nm obtained near \Tromso
 on February 17, 1996, from 20:29.30 to 20:34.40~UT. The spatial scale of
 each plot is $520\times520$~km$^2$. The white point near the middle of each
 plot indicates the location of \Tromso. The plot time is $10$~s. The
 \Tromso heater was turned on from 20:30 to 20:34~UT.
 }
 \label{plate:DASI_Feb17}
\end{plate*}

\begin{figure*}[p]
 \figurewidth{\linewidth}
 \figbox*{}{}{%
  \includegraphics[width=1.0\linewidth]{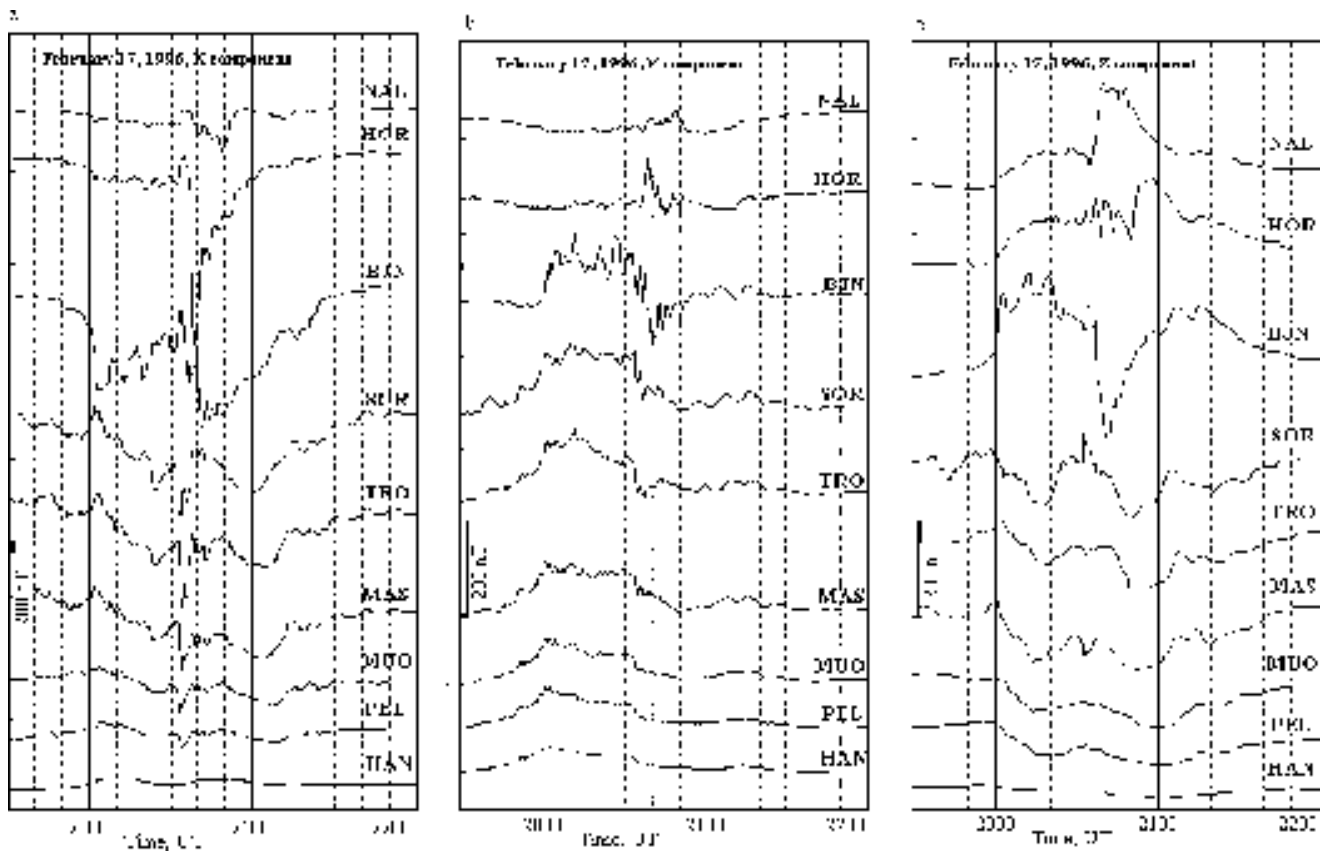}
 }
 \caption{
 The temporal behavior of the $X$, $Y$, and $Z$ components of the magnetic
 field variations on February 17, 1996. Data from the IMAGE magnetometers.
 }
 \label{fig:IMAGE_Feb17}
\end{figure*}

\section{Observational results}
\label{sec:results}

The \Tromso dynasonde ionograms as well as the altitude-temporal
variations of the electron density measured by the EISCAT UHF radar in
course of the \Tromso pumping experiments on February~16 and 17, 1996,
show the presence of an intense sporadic $\Es$ layer with a maximum
plasma frequency of $4.1\le{}f_0\Es\le4.5$~MHz at heights
$100\le\hm\Es\le110$~km.  We therefore conclude that the $E$ region of
the auroral ionosphere was actually the region where the powerful HF
radio waves were reflected.  This led to the generation of artificial
field-aligned irregularities (AFAIs) which scattered the $9$ and
$12$~MHz signals used for Doppler diagnostics.  It is believed
\citep{Djuth&al85,Noble&al87} that a thermal resonance instability at
the upper-hybrid (UH) level is the strongest candidate for the
excitation of AFAIs in the $E$ region.  It should be noted that, under
the specific conditions when the pump waves are reflected from auroral
$\Es$ layers, the pump reflection height is close to the altitude of the
heater-enhanced conductivity region which coincides with the auroral
electrojet; see the review article by \citet{Stubbe96} and references
therein.  This means that the plasma resonance level was close to the
altitude of the heater-enhanced conductivity region.

\subsection{Experiment on February 17, 1996}

The experiment on February~17, 1996, was conducted from 20 to 23~UT with
a 4~min on, 6~min off HF cycle.  A very interesting observation in this
experiment is the behavior of the auroral arc in the vicinity of \Tromso
during the heating cycle 20:30--20:34~UT; see the DASI data in
Plate~\ref{plate:DASI_Feb17}.  Beginning at 20:32~UT (third row in
Plate~\ref{plate:DASI_Feb17}), a gradual thinning of the auroral arc,
accompanied by the appearance of a weak bulge above \Tromso, is
observed.  Thereafter the brightening and subsequent breakup of the arc
20:33.40~UT takes place exactly above \Tromso.

The magnetic field $X$, $Y$, and $Z$ components, as recorded by the NAL,
HOR, BJN, SOR, TRO, MAS, MUO, PEL, and HAN stations of the IMAGE
magnetometer network, are displayed in Figure~\ref{fig:IMAGE_Feb17}; \cf
also Figure~\ref{fig:geometry}.  It can be seen that a substorm
activation started at 20:00~UT.  A second activation started at
20:33~UT, indicated by a large negative spike in the magnetic $X$
component, localized in a narrow latitudinal region around \Tromso, from
the SOR to the MAS station.

Examination of the peculiarities in the behavior of the $X$ and $Z$
magnetic components in Figure~\ref{fig:IMAGE_Feb17} shows that a new
westward electrojet appeared at 20:33~UT exactly above \Tromso (reversal
of the $Z$ component from positive values at SOR to negative values at
MAS, with $Z\approx0$ at the TRO station), and maximal negative amplitude
of the $X$ component ($\Xmax=-130$~nT at \Tromso).  Thereafter, at
20:38~UT, a large substorm started at higher latitudes ($\Xmax=-400$~nT
at the HOR station).  We will only consider the substorm activation
around \Tromso which started at 20:33~UT.

\begin{figure}{!t}
 \figurewidth{\linewidth}
 \figbox*{}{}{%
  \includegraphics[width=1.0\linewidth]{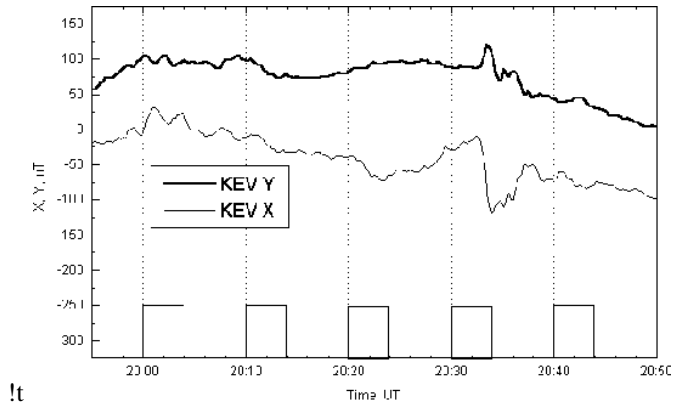}
 }
 \caption{
 Magnetic $H$ and $D$ components measured at the Kevo (KEV) magnetic
 station on February 17, 1996, from 19:55 to 20:50~UT. The heater-on
 periods are indicated on the bottom axis.
 }
 \label{fig:Kevo_Feb17}
\end{figure}

\begin{figure}{!t}
 \figurewidth{\linewidth}
 \figbox*{}{}{%
  \includegraphics[clip,viewport=120 150 710 560,width=1.0\linewidth]
   {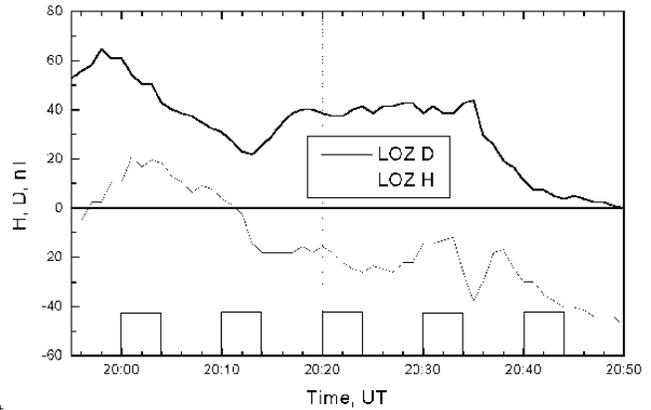}
 }
 \caption{
 Magnetic $X$, $Y$, $Z$ components measured at the Lovozero (LOZ) magnetic
 station on February 17, 1996, from 19:55 to 22:50~UT. The heater-on
 periods are indicated on the bottom axis.
 }
 \label{fig:Lovozero_Feb17}
\end{figure}

Studies of magnetograms from the Kevo (Figure~\ref{fig:Kevo_Feb17}) and
Lovozero (Figure~\ref{fig:Lovozero_Feb17}) stations, which are both
located further east of \Tromso, clearly show that an intensity of the
magnetic disturbance is decreasing with distance from \Tromso.  Hence,
we can safely conclude that the substorm really occurred above \Tromso.

\begin{figure}
 \figurewidth{\linewidth}
 \figbox*{}{}{%
  \includegraphics[width=1.0\linewidth]{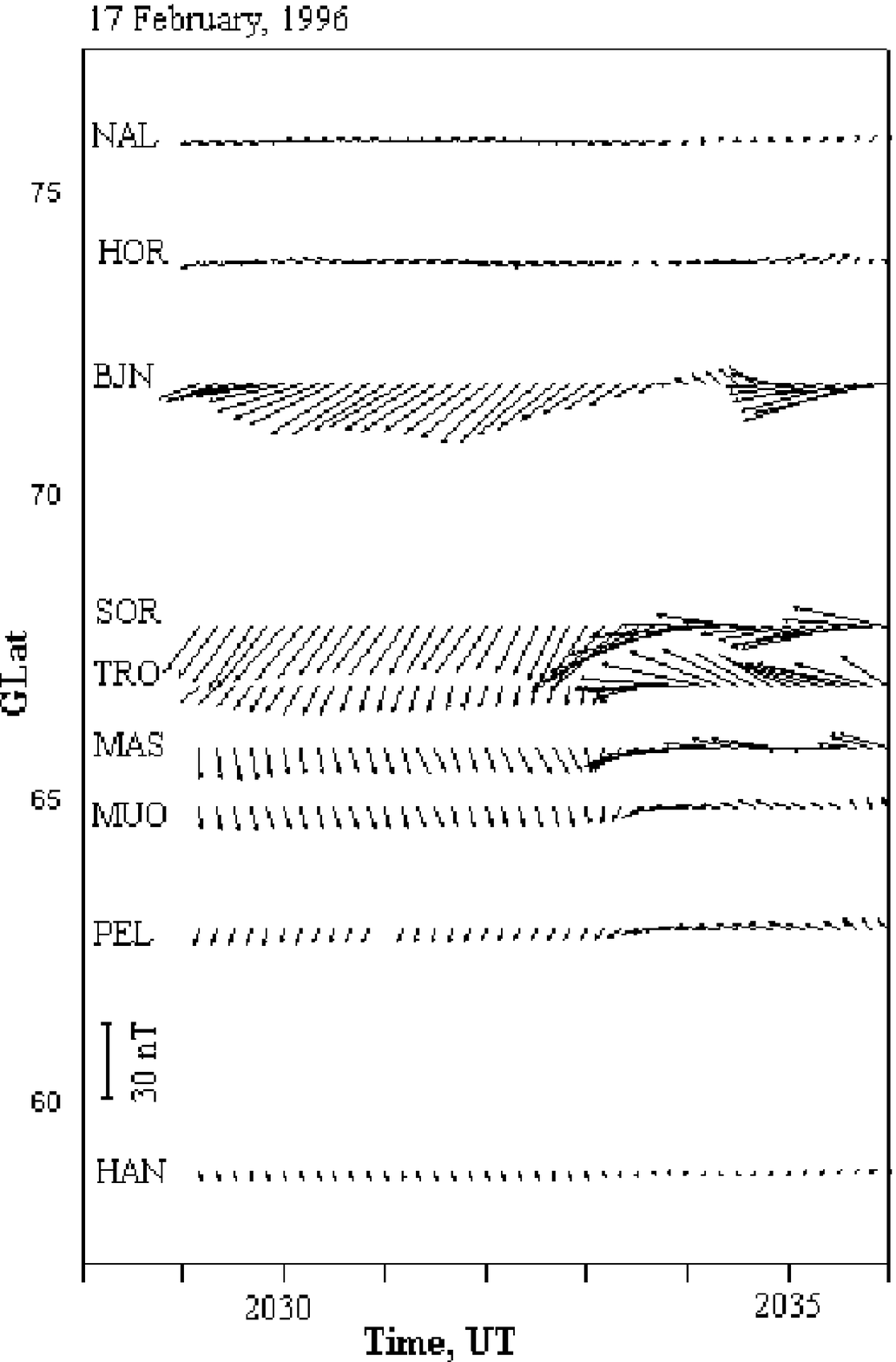}
 }
 \caption{
 The behavior of the equivalent current vectors on February 17, 1996,
 from 20:29 to 20:36~UT, describing the distribution of magnetic disturbances
 at the IMAGE network. The heater-on period was from 20:30 to 20:34~UT.
 }
 \label{fig:currvec_Feb17}
\end{figure}

The behavior of the equivalent current vectors (from the $X$ and $Y$
magnetic components in Figure~\ref{fig:IMAGE_Feb17}), describing the
distribution of the magnetic disturbances from 20:29 to 20:36~UT, is
summarized in Figure~\ref{fig:currvec_Feb17}.  It is clearly seen that
the most drastic changes of the current directions and magnitudes in the
vicinity of \Tromso from SOR to MAS occurred during the period
20:33-20:34~UT.

\begin{figure}[!t]
 \figurewidth{\linewidth}
 \figbox*{}{}{%
  \includegraphics[clip,viewport=80 150 400 760,width=1.0\linewidth]
  {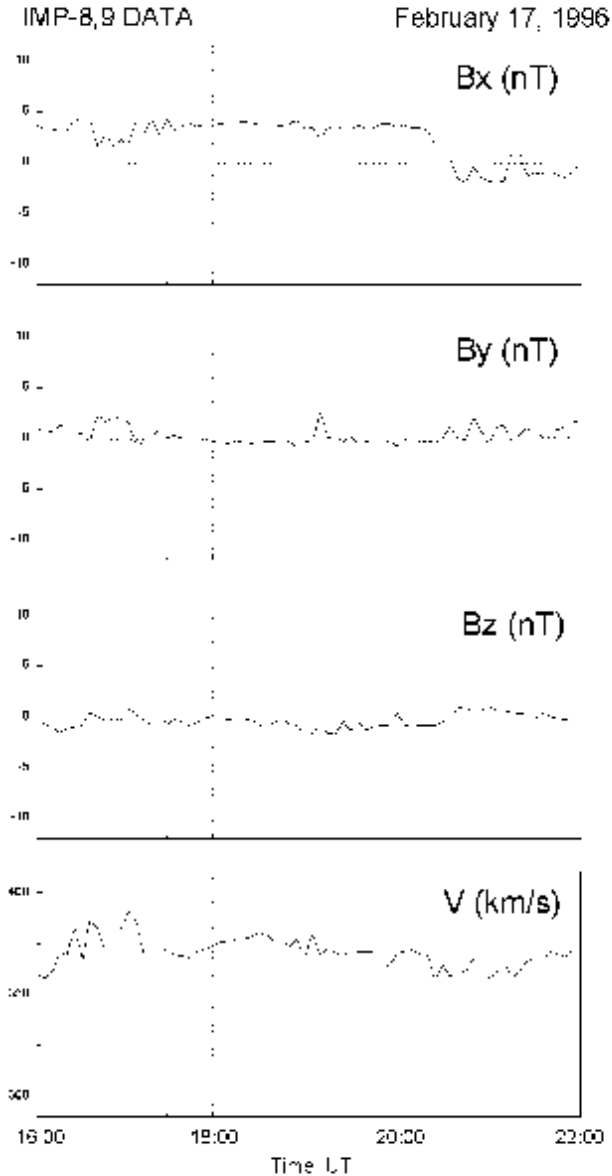}
 }
 \caption{
The behavior of the $\Bx$, $\By$, and $\Bz$ components of the
interplanetary magnetic field, and the velocity of the
solar wind, on February 17, 1996, from 16 to 22~UT.
IMP\,8 and IMP\,9 satellite data.
 }
 \label{fig:IMF_Feb17}
\end{figure}

It is well established that the onset of a substorm can be related to
disturbances in the interplanetary magnetic field (IMF) and/or in the
solar wind.  Because of that it is of interest to consider the IMF and
solar wind data during heating experiment.  The behavior of $\Bx$,
$\By$, and $\Bz$ components of the IMF as well as the solar wind
velocity $V$ from IMP\,8 and IMP\,9 atellite data data is shown in
Figure~\ref{fig:IMF_Feb17}.

As is clearly seen from Figure~\ref{fig:IMF_Feb17}, there are no
significant changes in the $\Bz$ and $\By$ components; during the period
16--22~UT they undergo maximal amplitude fluctuations not exceeding
1.5~nT.  Nonetheless, it should be pointed out that small directional
changes in the $\Bz$ component (northward turning) accompanied by a
$\Bx$ turn from positive to negative values occurred at about 20:35~UT.
Therefore, IMF data show an absence of major directional changes in the
$\Bz$ and $\By$ components which could be associated with the substorm
activation at 20:33:40~UT.  In this respect, we note the results of
detailed observations performed by \citet{Henderson&al96} and which
clearly show that the magnetospheric substorm can indeed occur in the
absence of of an identifiable driver in either the IMF or the solar wind
dynamic pressure.  It is suggested that the internal instability in the
magnetospheric system could be the possible driver for these substorms.

\begin{figure}[!t]
 \figurewidth{\linewidth}
 \figbox*{}{}{%
  \includegraphics[width=1.0\linewidth]{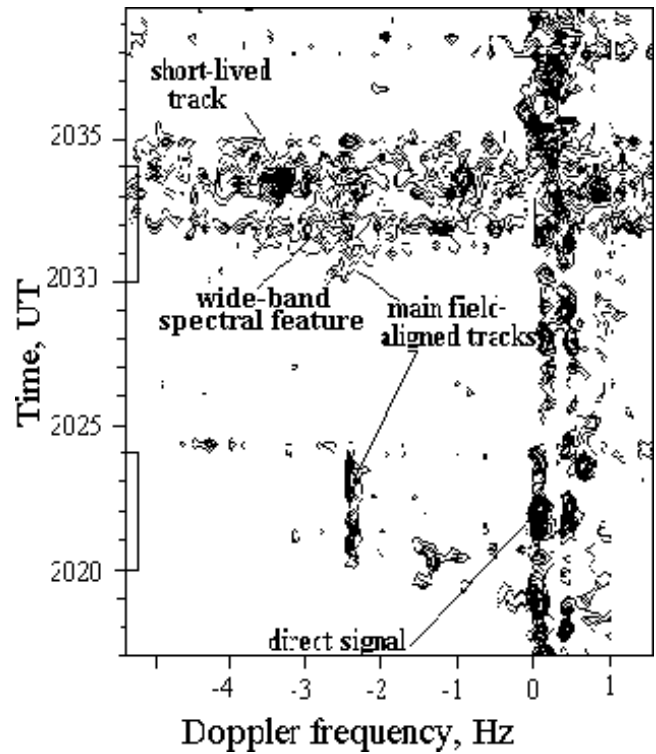}
 }
 \caption{
 Dynamic Doppler spectra of HF diagnostic signals on the
 London--\Tromso--St.~Petersburg path at the operational frequency $12,095$~kHz
 on February 17, 1996, from 20:18 to 20:38~UT. The direct signals propagating
 from the transmitter to the receiver along a great circle path correspond
 to zero Doppler shifts. The intervals when the \Tromso heater was
 turned on are marked on the time axis.
 }
 \label{fig:dyn_doppler_Feb17}
\end{figure}

Figure~\ref{fig:dyn_doppler_Feb17} presents dynamic Doppler spectra of
HF diagnostic signals recorded in St.~Petersburg in the course of two
heating cycles from 20:18 to 20:38~UT.  The turning on of the \Tromso
heater at 20:20~UT led to the appearance of an additional track, shifted
from the direct signal by about $-2.5$~Hz.  This additional track
disappeared after the heater was turned off at 20:24~UT and was produced
by diagnostic waves scattered from the artificial field-aligned
irregularities (AFAIs) generated by the \Tromso HF heating facility in
the auroral $E$ region.

As can also be seen from Figure~\ref{fig:dyn_doppler_Feb17}, the heater
turn-on at 20:30~UT led to the appearance of field-aligned scattered HF
signals in the same manner as in the preceding heating cycle.  About
$60$~s later, wide-band spectral feature occurred throughout the spectral
bandwidth analyzed.  Thereafter, at 20:33~UT, an additional very
intense, short-lived track of about 1~min duration, displaced $-0.8$~Hz
from the main scattered signal, appeared in the Doppler sonogram.  The
heater turn-off at 20:34~UT was followed by the disappearance of the
additional short-lived track, but the AFAI scattered signals, as well as
the wide-band spectral features, were maintained for yet another minute.

What is the nature of the wide-band spectral feature closely linked to
the auroral activation?  It is known that one of the most common
features in HF pumping experiments is the generation of Stimulated
Electromagnetic Emissions (SEE)
\citep{Thide&al82,Thide&al83,Thide&al89,Thide90, Stubbe&al84,
Leyser&al94}.  Unfortunately, no SEE diagnostics capability was
available during the experiments so it was not possible to establish
directly whether the basic SEE component was excited or not.

The long-delayed effect of about one minute duration, of the wide-band
feature observed after the heater turn-off is not clear.  Because of
this we do not exclude the possibility that this heater-related emission
in our experiment is a heater-modified natural auroral emission in the
decameter range \citep{LaBelle&al95,Weatherwax&al95}.

Another possible explanation is that the observed heater-related
wide-band features are accompanied by the excitation of VLF waves and
turbulence, known to be excited in ionospheric HF pumping.  As was
concluded by \citet{Vaskov&al98} from satellite experiments during the
action of un-modulated powerful HF radio waves on the nightside
ionospheric $F$ region, VLF waves may be excited due to a decay process
taking place near the pump wave reflection region, or due to the
interaction with suprathermal electrons, accelerated by the
pump-enhanced plasma turbulence.  The heater induced VLF waves can be
studied by ground-based techniques.  On the other hand, HF-excited VLF
waves in the whistler mode were detected by satellites in the upper
ionosphere and magnetosphere within the magnetic flux tube footprinted
on the heating facility \citep{Vaskov&al98}.

Closely related to the wide-band spectral features in the probe wave
sidebands was the appearance of an additional short-lived track in the
Doppler sonogram.  It is likely that this short-lived track was induced
by a stimulated precipitation of electrons due to a cyclotron resonant
interaction of natural precipitating electrons with heater-induced
whistler waves in the magnetosphere
\citep{Trefall&al75,Bosinger&al96,Trakhtengerts99}.  From the results
obtained, we may then conclude that the substorm activation on
February~17, 1996, at 20:33~UT exactly above \Tromso was initiated by
the pump-induced electron precipitation.

\begin{figure*}[p]
 \figurewidth{\linewidth}
 \figbox*{}{}{%
  \includegraphics[width=0.8\linewidth]{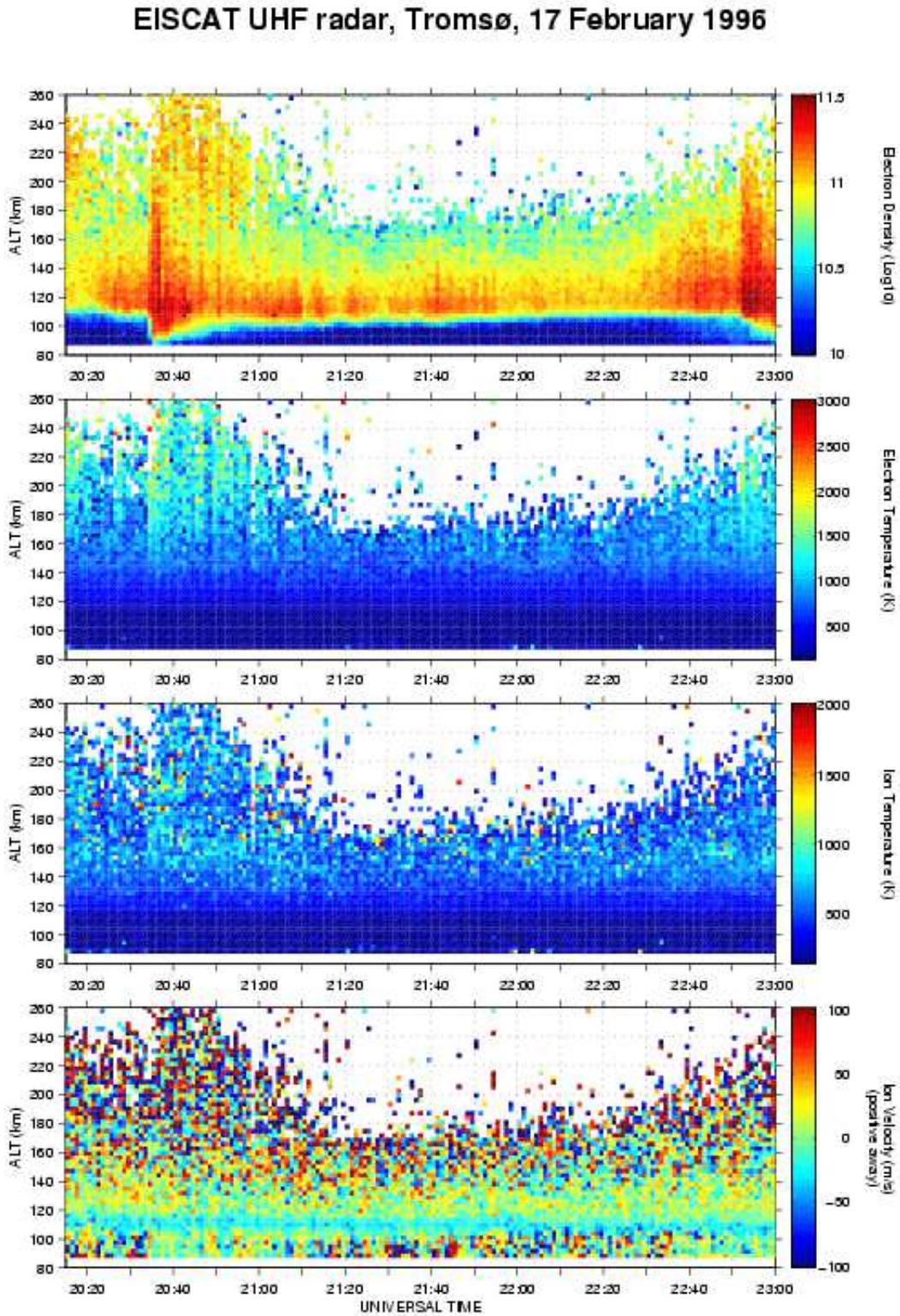}
 }
 \caption{
 Behavior of the electron density, electron and ion temperatures,
 and ion velocity as observed with the \Tromso EISCAT UHF incoherent
 scatter radar February~17, 1996, from 20:15 to 23:00~UT, by using a high
 spatial resolution alternating code (AC). The \Tromso HF heating facility was
 operated from 20:20 to 23~UT with a 4~min on, 6~min off pump cycle.
 }
 \label{fig:EISCAT_Feb17}
\end{figure*}

Another interesting peculiarity was detected in the EISCAT UHF radar
data (Figure~\ref{fig:EISCAT_Feb17}).  The formation of the electron
density cavity took place at 20:33~UT downward from the altitude of
about $100$~km.  It was accompanied by the occurrence of a burst-like
increase of the electron density and temperature, $\Ne$ and $\Te$, in a
wide range of altitudes upward from $110$~km.  It is well known that an HF
pump wave may excite plasma waves and turbulence in the resonant region
where the ordinary mode ($O$-mode) of the pump wave is reflected from
the ionosphere and in the upper hybrid, UH, resonance region.  The
threshold field strength for the generation, due to ponderomotive force
effects, of plasma waves by a powerful $O$-mode radio wave is of the
order $\Eth\approx200$~mV/m \citep{Fejer79,Thide90}, a value which was
significantly exceeded in our experiments.  As was pointed out by
\citet{Fejer79}, partial pressure effects can lower this threshold.  In
these processes, the electrons are assumed to be accelerated and this
leads to an electron flux transport along the magnetic field lines
\citep{Bernhardt&al88}.  Note that the observed changes in the $\Ne$ and
$\Te$ are closely correlated with the auroral activation and may
therefore be a signature of the heater-induced precipitation of
electrons.

Summarizing the experimental findings from the different ground-based
measurements during the heater-on period 20:30--20:34~UT on February 17,
one can distinguish the following peculiarities related to the auroral
activation observed after 30 minutes from the start of the pumping
experiment in the absence of the apparent drivers in the IMF/solar wind
parameters:
(a) modification of the auroral arc and its break-up above \Tromso;
(b) local changes of the horizontal currents in the $E$ region;
(c) generation of scattered Doppler components throughout the whole spectral
bandwidth of 33~Hz analyzed;
(d) appearance of an additional short-lived Doppler sonogram track, distinct
from the main track, corresponding to scattered diagnostic signals due to
pump-induced electron precipitation;
(e) formation of the cavity in $\Ne$ downward from $100$~km and burst-like
increase of $\Ne$ and $\Te$ at heights upward from $110$~km;
(f) substorm activation exactly above \Tromso.

\begin{plate*}[p]
 \figurewidth{\linewidth}
 \figbox*{}{}{%
  \includegraphics[width=1.0\linewidth]{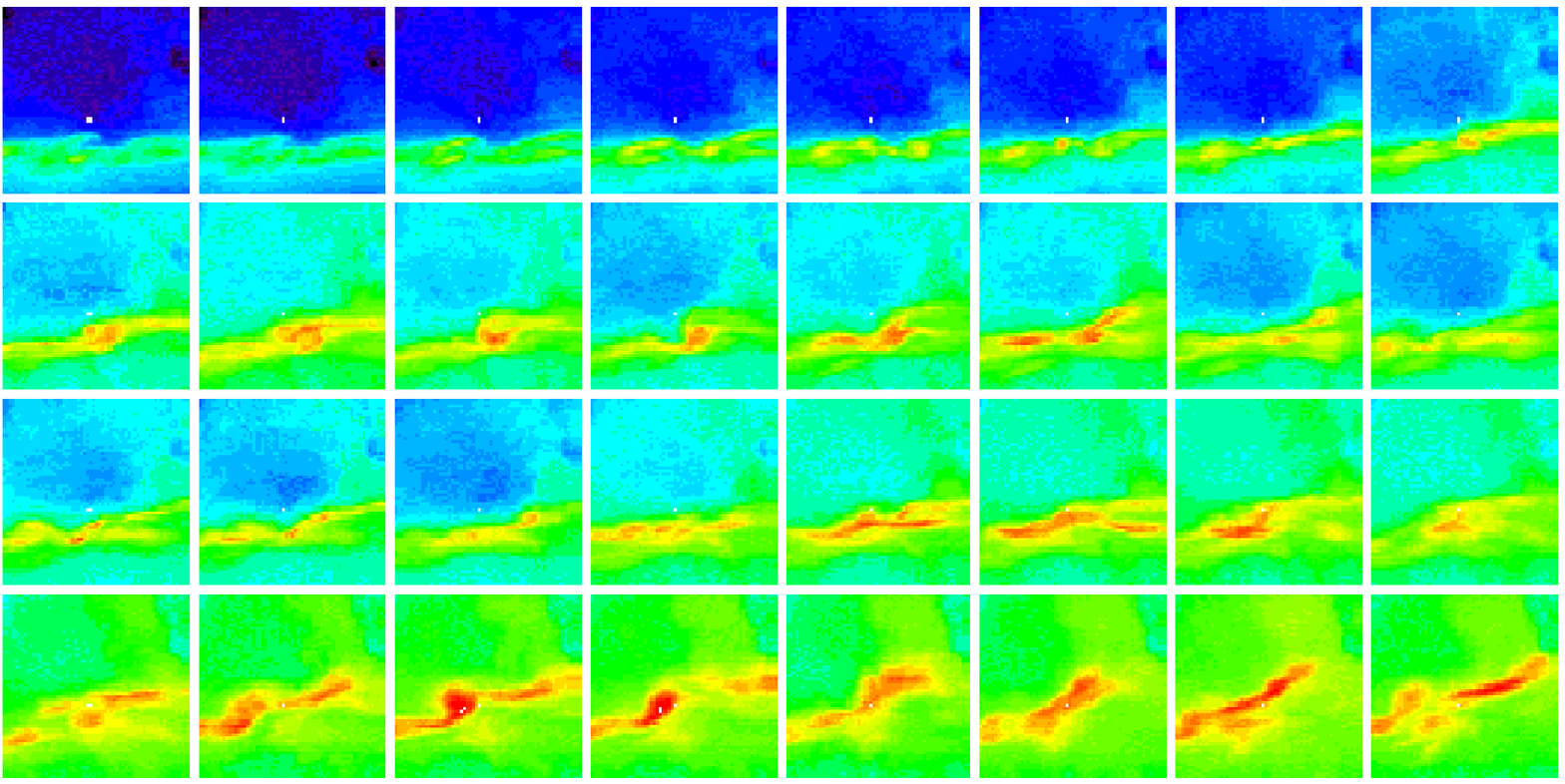}
 }
 \caption{
 A sequence of all-sky imager plots at $557.7$~nm obtained near \Tromso
 on February 16, 1996, from 21:18:20 to 21:33:50~UT. The spatial scale of
 each plot is the same as in Figure~\ref{plate:DASI_Feb17}.  The plot
 time is $30$~s. The \Tromso heater was turned on from 21:20 to 21:24~UT
 and 21:30 to 21:34~UT.
 }
 \label{plate:DASI_Feb16}
\end{plate*}

\begin{figure*}[p]
 \figurewidth{\linewidth}
 \figbox*{}{}{%
  \includegraphics[width=1.0\linewidth]{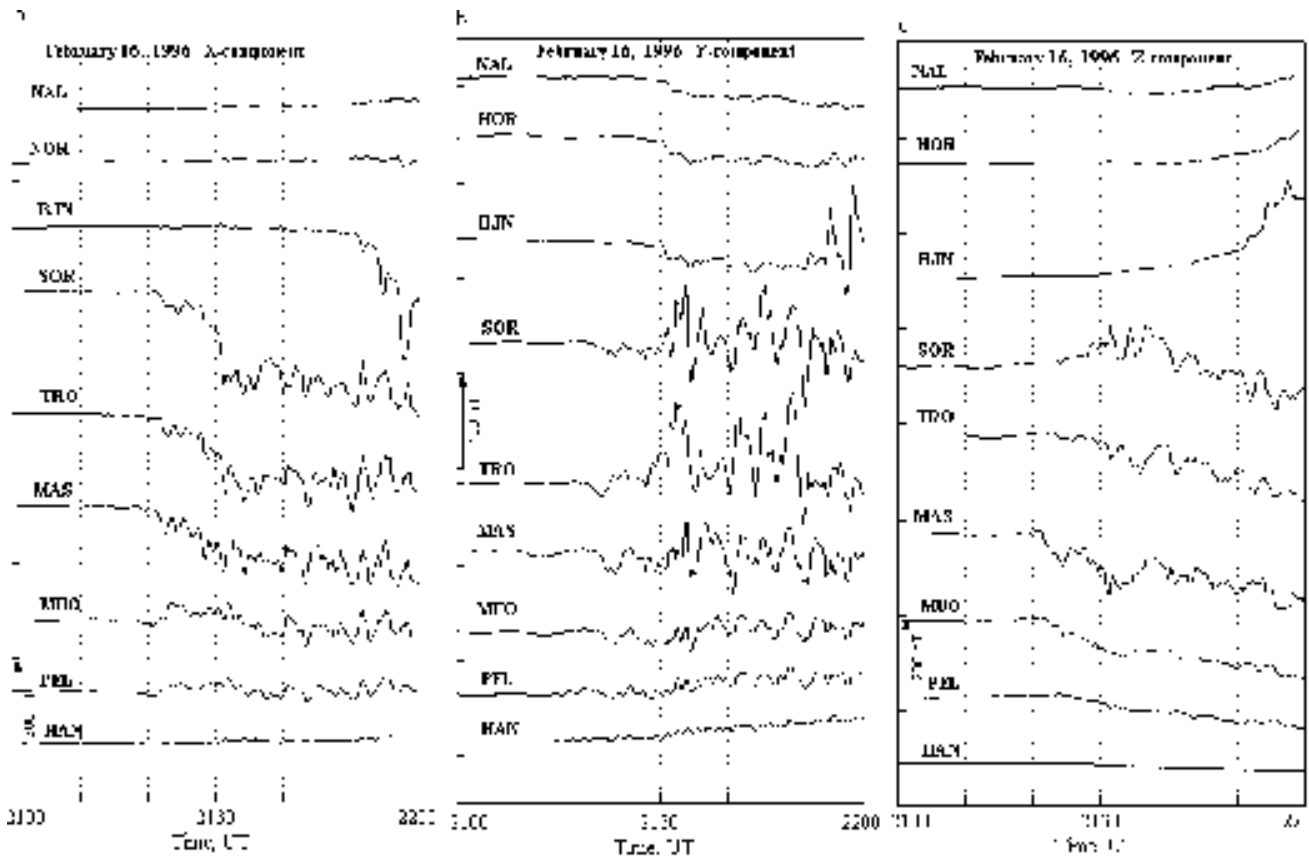}
 }
 \caption{
 The temporal behavior of the $X$, $Y$, and $Z$ components of the magnetic
 field variations on February 16, 1996. Data from the IMAGE magnetometers.
 }
 \label{fig:IMAGE_Feb16}
\end{figure*}

\subsection{Experiment on February 16, 1996}

In this section we present data from the \Tromso HF pumping experiment
carried out on February 16, 1996, starting at 21:00~UT.  The pump wave
was modulated with a 4~min on, 6~min off cycle from 21 to 23~UT.  This
was preceded by some short on periods as the HF transmitter was tuned,
from 20:41 to 20:44~UT, from 20:46:30 to 20:48:30~UT, and from 20:52:00
to 20:52:20~UT.  The DASI digital all-sky imager data for this day
(Plate~\ref{plate:DASI_Feb16}) show that the latitude-oriented auroral
arc was located slightly to the south of the heater.  It should be
pointed out that a most remarkable optical phenomenon was observed
during the two heating cycles 21:20--21:24~UT and 21:30--21:34~UT.  As
can be seen in Plate~\ref{plate:DASI_Feb16}, a development of local
spiral-like forms in the auroral arc near \Tromso occurred after the
heater was turned on.  In the first case (21:20--21:24~UT), the spiral
form appeared at 21:21:50~UT and led to the start of an auroral
activation.  In course of the heater-on period the intensity of this
spiral increased and started to decay only after the heater was turned
off.  In the second case (21:30--21:34~UT), a similar form, but more
intense in comparison with the heater-on period 21:20--21:24~UT,
appeared at 21:31:20~UT.  Furthermore, the brightening and subsequent
break-up of an auroral arc at 21:33:50~UT (Plate~\ref{plate:DASI_Feb16},
fourth row, last panel) took place exactly above \Tromso.  Such a spiral
form can be attributed to the local appearance of field-aligned currents
during the heater-on periods 21:20--21:24~UT and 21:30--21:34~UT.

\begin{figure}[!t]
 \figurewidth{\linewidth}
 \figbox*{}{}{%
  \includegraphics[width=1.0\linewidth]{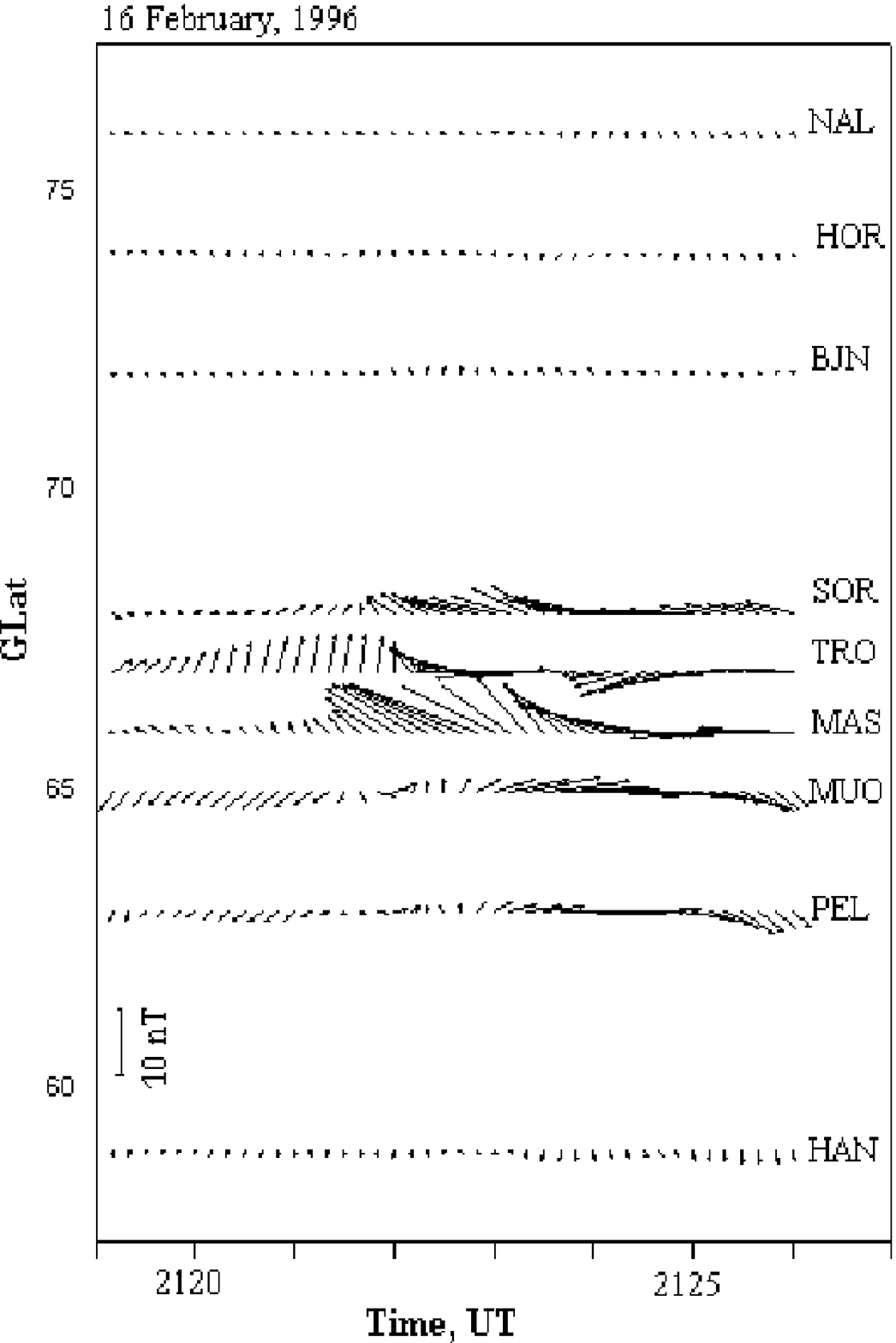}
 }
 \caption{
 The behavior of the equivalent current vectors on February 16, 1996,
 from 21:19 to 21:27~UT, describing the distribution of magnetic disturbances
 at the IMAGE network. The heater-on period was from 21:20 to 21:24~UT.
 }
 \label{fig:currvec_Feb16}
\end{figure}

IMAGE magnetograms for the event on February 16, 1996 ($X$, $Y$, and $Z$
magnetic components) are displayed in Figure~\ref{fig:IMAGE_Feb16}.
These magnetic data indicate that the \Tromso heater operation started
under quiet magnetic conditions (from 21:00~UT) and that the start of
the auroral activation occurred at 21:21~UT during the third heating
cycle (21:20--21:24~UT).  It should be noted that this event, just as
the substorm activation on February 17, 1996, occurred in a narrow
latitudinal region localized around \Tromso.  This is evident from the
behavior of the equivalent current vectors obtained from the IMAGE
magnetometers from 21:19 to 21:27~UT (Figure~\ref{fig:currvec_Feb16}).
Moreover, the most drastic changes of the current directions and
magnitudes were also observed in the vicinity of \Tromso, ranging in
latitude from the SOR to the MAS magnetic stations.

\begin{figure}[!t]
 \figurewidth{\linewidth}
 \figbox*{}{}{%
  \includegraphics[clip,viewport=70 150 390 760,width=1.0\linewidth]
  {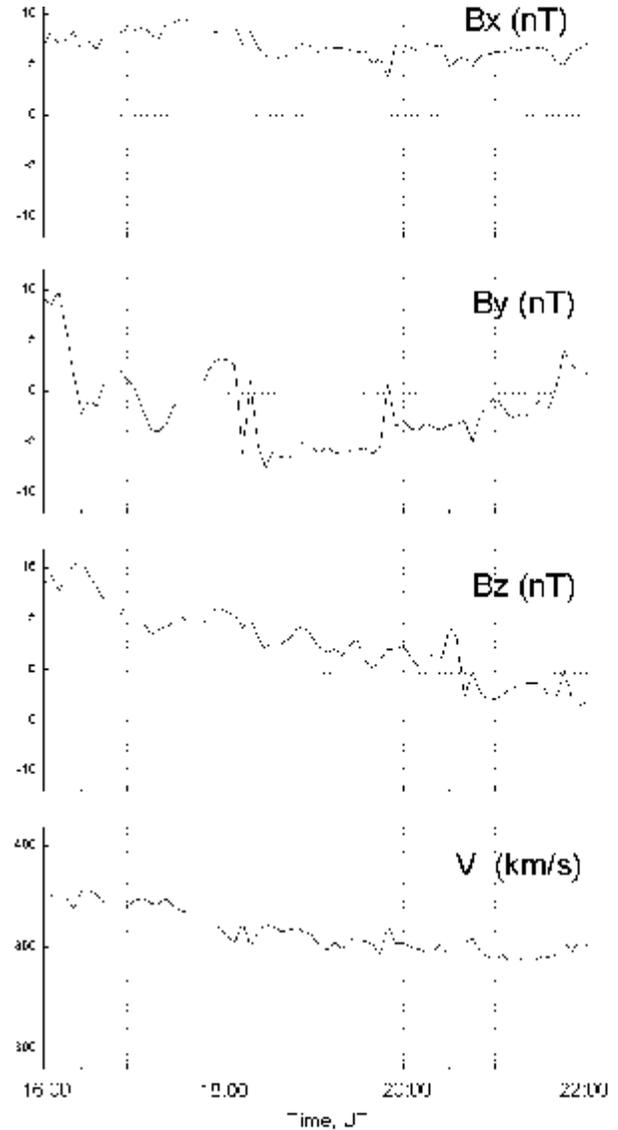}
 }
 \caption{
  The same as in Figure~\ref{fig:IMF_Feb17} but for February 16, 1996.
 }
 \label{fig:IMF_Feb16}
\end{figure}

The $\Bx$, $\By$, and $\Bz$ components of the IMF and the solar wind
velocity obtained from IMP\,8 and IMP\,9 satellite data for the event on
February~16, 1996 are shown in Figure~\ref{fig:IMF_Feb16}.  Contrary to
the event on February~17, a southward turning of the $\Bz$ component of the
IMF took place at 20:40~UT.  The amplitude of the southward $\Bz$
component was about of $-2$~nT.  This small directional change in the
$\Bz$ component could possibly be a driver for the auroral activation.

\begin{figure}[!t]
 \figurewidth{\linewidth}
 \figbox*{}{}{%
  \includegraphics[width=1.0\linewidth]{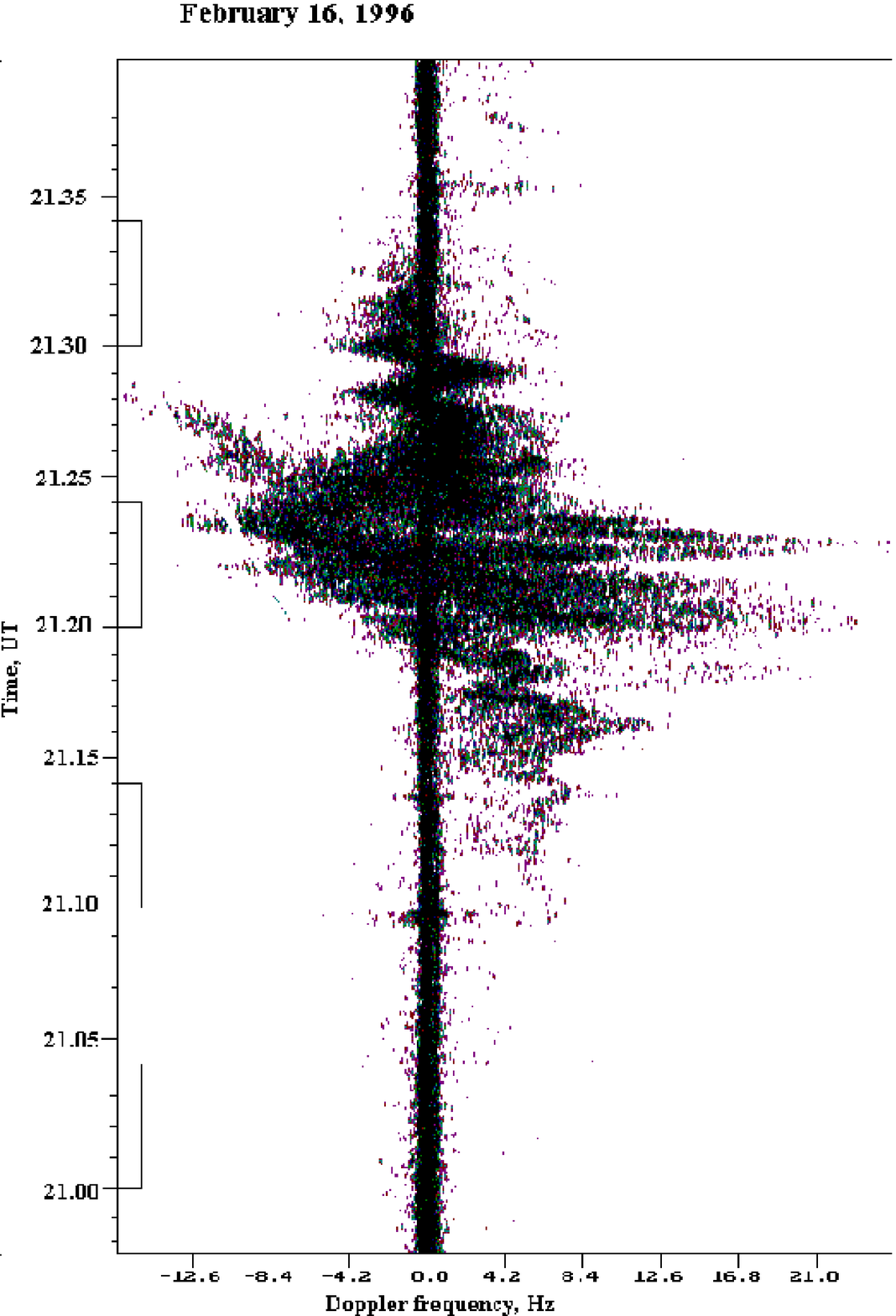}
 }
 \caption{
 Dynamic Doppler spectra of HF diagnostic signals on the
 London--\Tromso--St.~Petersburg path at operational frequency
 $9410$~kHz on February 16, 1996, from 20:58 to 21:40~UT.  The intervals
 when the \Tromso heater was turned on are marked on the time axis.
 }
 \label{fig:dyn_doppler_Feb16}
\end{figure}

Figure~\ref{fig:dyn_doppler_Feb16} displays the dynamic Doppler spectra
obtained on February~16, 1996, from 20:58 to 21:40~UT, on the
London--\Tromso--St.~Petersburg path, for a radio scatter operational
frequency of $f=9410$~kHz.  One can see that the heater turn-on at
21:10~UT led to the appearance of a weak scattered signal shifted from
the direct signal (corresponding to zero Doppler shift) by about
$+6.3$~Hz.  Note that the Doppler frequency $\fd$ of the scattered
signal changed during this heater-on period with a maximum magnitude of
$2.2$~Hz.  When the heater was turned off at 21:14~UT, the scattered
signals did not disappear as was the case on February 17, 1996.  Up to
21:34~UT, intense scattered signals from natural $\Es$ irregularities
were observed in St.~Petersburg.

The spectral structure of the signals scattered from AFAIs is quite
complicated.  It includes a broadening of the Doppler spectra and
burst-like noise enhancements.  The variations of $\fd$ with time of the
broad part in the dynamic Doppler spectra can be correlated with the
movements of the auroral arcs in the vicinity of \Tromso on the line of
sight of the radio scatter observations from St.~Petersburg.

The noise enhancement occurred over a frequency
range of up to $35$~Hz, which is below the ion cyclotron frequency.  We
emphasize that the Doppler measurements in this modification experiment
were performed in a limited frequency bandwidth of $50$~Hz.  At
frequencies below the ion cyclotron frequency the only known
electromagnetic modes of propagation along magnetic field lines is the
Alfv\'en wave.  Thus one would expect that Alfv\'en waves
would be excited by the HF pumping of the night-side auroral $E$
region.  Alfv\'en waves associated with ELF noise are identified with
electric to magnetic field ratios of the order of the Alfv\'en velocity.
These waves are observed to occur in narrow regions typically of the
order $1$--$3$~km and have highly irregular wave forms
\citep{Gurnett&al84}.

\begin{figure*}[p]
 \figurewidth{\linewidth}
 \figbox*{}{}{%
  \includegraphics[width=0.8\linewidth]{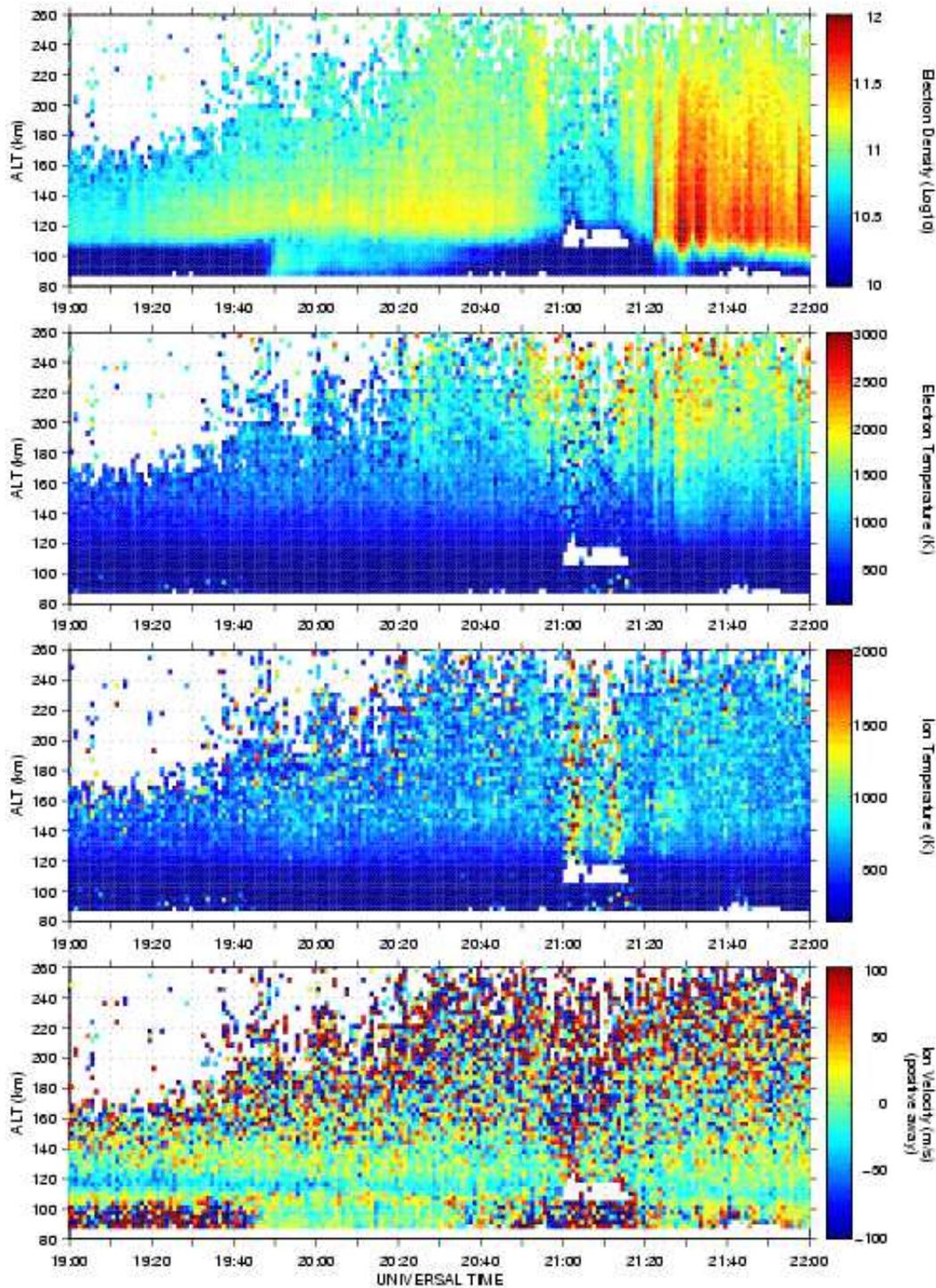}
 }
 \caption{
Behavior of the electron density, electron and ion temperatures, and
ion velocity from \Tromso EISCAT UHF radar measurements on February 16,
1996, from 19 to 22 UT obtained by the use of the high spatial resolution
alternating code (AC). The interval from 21 to 21:15~UT with a "missing
data layer" at about $110$~km is almost certanly data which could not be
fitted by the standard program because of enhanced electron temperatures
produced by the Farley-Buneman instability due to the large drifts
(horizontal electric fields). The enhanced ion temperature over the whole
$120$--$260$~km height range is evidence of this enhanced fields. All these
features are the recognized phenomena. The \Tromso HF heating facility
was operated from 21 to 22~UT with a 4 min on, 6 min off pump cycle.
 }
 \label{fig:EISCAT_Feb16}
\end{figure*}

Figure~\ref{fig:EISCAT_Feb16} presents EISCAT \Tromso UHF radar data
($\Ne$, $\Te$, $\Ti$, and $\Vi$) obtained on February~16, 1996, from 19
to 22~UT.  It can be seen from Figure~\ref{fig:EISCAT_Feb16} that in the
first two consecutive heating cycles from 21:00 to 21:04~UT and 21:10 to
21:14~UT, increases of $\Ne$ and $\Te$ in the altitude range
$120$--$160$~km were clearly observed.  Recall that the HF pump wave was
reflected from heights of about $100$--$110$~km.  Note also the ``missing
data layer'' at about $110$~km from 21 to 21:15~UT.  This is almost
certainly due to data which could not be fitted by the standard program
because of enhanced electron temperatures due to a Farley-Buneman
instability produced by large drifts of horizontal electric fields.
The enhanced ion temperature $\Ti$, over the whole $120$--$160$~km height
range is an evidence of this enhanced field.  All these features are
recognized phenomena.  Therefore, the start of the \Tromso HF heater
operation at 21~UT was accompanied by the strong Farley-Buneman
instability excitation at the HF pump wave reflection level.

Similar to the event of February~17, 1996, we observed in the EISCAT UHF
radar data during the auroral activation onset at 21:21~UT a formation
of an electron density cavity downward from the height of $110$~km.  This
was accompanied by a burst-like increase of $\Ne$ at altitudes upward
from $110$~km.  Furthemore, once started at about 21:21~UT, the process of
the auroral activation development was able maintain itself even when
the heater was turned off.  This can be seen from
Figure~\ref{fig:EISCAT_Feb16} which exhibits electron density
increases before 21:30~UT, at which time the HF heater was turned off.

From the multi-instument experimental data from the period
21:20--21:24~UT on February~16, 1996, one can identify the following
specific features, which may be related to the auroral activation
observed 20~minutes from the start of the HF heater operation in
the presence of a small directional change of the $\Bz$ component of the
IMF:  (a) modification of the auroral arc and local spiral-like
formation; (b) local changes of the horizontal currents in the $E$
region in the vicinity of \Tromso; (c) generation of burst-like noise in
the frequency range up to $35$~Hz; (d) formation of the cavity in $\Ne$
downward from a $110$~km level and burst-like increase in $\Ne$ at heights
upward from $110$~km; (e) substorm activation in the localized
latitudinal region above \Tromso.

\section{Discussion}
\label{sec:discussion}

A bistatic HF Doppler radio scattering setup has been used in
conjunction with the IMAGE magnetometer network, the DASI digital
all-sky imager, the \Tromso dynasonde, and the EISCAT UHF radar to find
evidence that powerful HF radio waves can cause a modification of the
ionosphere-magnetosphere coupling which can lead to a local
intensification of the auroral activity.  Results presented, as obtained
from ground-based observations made on two consecutive days, can be
interpreted as auroral activations localized over the \Tromso HF heating
facility.  In recent times, the possibility of triggering magnetospheric
substorms by artificial impacts has been discussed.  The artificial
localization of a magnetospheric substorm by strong HF radio beams was
considered by \citet{Mogilevsky99}, based on measurements onboard the
INTER\-BALL-2 (Auroral probe) satellite.  From a consideration of
results of active experiments, \citet{Foster98} concluded that substorm
onsets may be initiated by such actions.  A distinctive feature of the
experiments reported here is that the HF pump waves were reflected from
an auroral $\Es$ layer.  The presence of a sporadic $\Es$ layer as
well a an auroral arc in the vicinity of the \Tromso heating facility are
indicative of naturally precipitating electrons with an energy of a few
keV.

First of all, it should be mentioned that there are two basic
possibilities to locally activate the auroral activity by an artificial
ionospheric modification.  The simplest one is creating such strong a
perturbation that the surrounding currents exceed the threshold for the
instability producing energetic electrons.  It implies that the
height-integrated conductivity, $\Sigma=\alpha\int{}\Ne(z)dz,$ be
changed by a factor of 5--10 inside a region of more than $10$~km
dimension.  This may probably happen during injection of charged
particle beams and plasma clouds in the ionosphere (described by
\citet{Holmgren&al80}).

\begin{sloppypar}
Another possibility initially explored by \citet{Zhulin&al78} is based
on the idea of a positive feedback in the magnetosphere-ionosphere
coupling during the substorm growth phase (see different scenarios in
\citep{Kan93,Trakhtengerts&Feldstein91,Lysak91}.  In this way, an
artificially created conductivity perturbation in the $E$ region grows
faster than in the vicinity until a critical value of the field-aligned
current (FAC) has been achieved.  Note, that in our \Tromso HF pumping
experiments, the auroral activations were observed after 20--30~minutes
from the start of the HF heater operation.  This is a significant
indication that a positive feedback mechanism in the
magnetosphere-ionosphere coupling is preferred over other mechanisms to
produce the local auroral activations by HF pump waves beamed into a
sporadic $\Es$ layer.
\end{sloppypar}

During the experiments, local changes of the ionospheric conductivities,
and therefore currents in the magnetic flux tube footprinted at \Tromso,
were observed.  Note that the duration of the HF heater period was 4~min
in our HF pumping experiments.  In this long-period heating cycle,
conductivity changes are produced due to the electron density
disturbances.  Their effect is much stronger than those of the
short-period heating cycles \citep{Stubbe96}.  The results of
calculations for long-period heating cycles have shown that the
conductivity perturbations in the HF pump-modified region are the order
of $\delta\Sigma(r)=2$~S \citep{Lyatsky&al96}.  Very drastic local
horizontal current changes, closely correlated with auroral activations
during the heater-on periods, were seen in the behavior of the
equivalent current vectors obtained from the $X$ and $Y$ magnetic field
components obtained by the IMAGE magnetometer network (see
Figures~\ref{fig:currvec_Feb17} and \ref{fig:currvec_Feb16}).  It is
quite possible that the region of the heater-enhanced ionospheric
conductivity was polarized in the background electric field and the
polarization electric field propagated into the magnetosphere along the
magnetic field lines as the field of an outgoing Alfv\'en wave
\citep{Lyatsky&Maltsev83,Kan&Sun85,Lysak90,Borisov&al96,Kozlovsky&Lyatsky97}.
However, other, non-linear processes may also have contributed to the
effects observed.

Another phenomenon which accompanied the auroral activations, was the
appearance of wide-band spectral features in the Doppler radio
scattering data.  There are grounds to assume that these spectral
features may be associated with pump-induced VLF and ELF noise.  Note
that these spectral features were observed when the \Tromso heating
facility was operated in an $4$~min on/$6$~min off mode, which is a
frequency which is significantly lower than that of the the waves
excited.  Therefore, the VLF and ELF noise in our experiments is not the
same as the well known effect of ELF and VLF excitation at combination
frequencies \citep{Getmantsev&al74,Rietveld&al89}.  The emissions
observed have other origin.  It should be pointed out that VLF and ELF
noise propagating into the topside ionosphere over large distances
excited by un-modulated HF pump waves were identified for the first time
by \citet{Vaskov&al98} in satellite experiments.

It is thought that on February~17, 1996, VLF waves at the harmonics of
the lower hybrid frequency were generated by the $150$~MW ERP HF waves
from the \Tromso heating facility.  These VLF waves propagated, in the
whistler mode, along the \Tromso magnetic field line into the upper
ionosphere.  Their interaction with natural precipitation electrons due
to a cyclotron resonance, led to a pump-induced precipitation of
electrons responsible for the occurrence of additional short-lived track
on the Doppler sonogram (see Figure~\ref{fig:dyn_doppler_Feb17}).  From
the results obtained we conclude that a substorm activation exactly
above \Tromso at 20:33~UT was initiated by the pump-induced
precipitation of the electrons.

One would expect that on February~16, 1996, Alfv\'en waves were excited
by the powerful HF radio waves beamed into nigt-side auroral E region.
These Alfv\'en waves can be associated with ELF noise enhancement
occurrring over a frequency range up to 35 Hz (see
Figure~\ref{fig:dyn_doppler_Feb16}).

It should be noted that the amplification of the interaction between the
magnetospheric convection and the ionospheric base is induced by
accelerated electrons.  What accelerator mechanisms can be caused by HF
pumping of the base of the ionosphere at $\sim100$~km?  It is known
\citep{Bernhardt&al88} that electrons are accelerated by the electric
potential associated with plasma waves excited by HF pump waves at the
plasma and upper hybrid resonance levels.  This leads to an electron
flux transport along the magnetic field lines.  Ohmic heating increases
the electron temperature $\Te$ and the plasma pressure in the modified
ionosphere.  An increase of the $\Te$ at heights above the HF pump
reflection level can be distinguished in the EISCAT UHF radar data (see
Figures~\ref{fig:EISCAT_Feb17} and \ref{fig:EISCAT_Feb16}).  HF pumping
may also lead to other possible accelerator mechanisms which occur in
the naturally disturbed auroral ionosphere:  lower hybrid wave Landau
resonance, kinetic Alfv\'en waves, and anomalous resistivity
\citep{Borovsky93,Vogt&Haerendel98}.

Let us consider a probable chain of events during the growth phase of a
substorm after a stable auroral arc has arrived, and the heater was turned
on.  Usually, a significant part of the upward current in the arc
current system $j_{z0}^\uparrow$ was carried by precipitating
energetic electrons that ionize the ionosphere.  In the positive
feedback instability scenario, perturbation of the plasma density and
that of the upward FAC intensity are tied as $\mathrm{d}\delta
\Ne/\mathrm{d}t\simeq{}Q\delta{}j_{z}^\uparrow$ [cm$^{-3}$s$^{-1}$].
Here $Q$ is the ionization function of the precipitating flux and
$j_{z}^\uparrow$ is in mA/m$^{2}$.

In turn, the perturbation of $j_{z}^\uparrow$ carried by an Alfv\'en
wave, is generated over a conductivity perturbation
$\delta\Sigma(\mathbf{r})$, can be approximated as (\eg,
\citep{Kozlovsky&Lyatsky97})
\begin{equation}\label{eq:jz}
 \delta j_{z}\simeq\frac{\SigmaA}{\SigmaP}
  \frac{\delta\Sigma}{{\SigmaP}_0}j_{z0}^\uparrow
 \simeq\beta j_{z0}^\uparrow\frac{\delta\Ne}{\Ne}
\end{equation}
where $\SigmaA=1/\mu_0\VA$ is the wave conductivity,
$\mu_0$ is the vacuum permeability, $\VA$ is Alfv\'en velocity,
$\SigmaP$ is the height-integrated Pedersen conductivity, and
$\beta\sim0.1$.  An initial inhomogeneity grows with a characteristic
rate (instability increment) of
$\gamma\simeq{}Q\beta{}j_{z0}^\uparrow/N_0\simeq10^{-3}$~s$^{-1}$ for
$Q=(3-10)10^2$, $N_0=10^5$~cm$^{-3}$, $j_{z0}^\uparrow=3$--$1$.  This
implies that waves propagate toward the source of energetic particles
without losses and their transit time is short, $\ll\gamma^{-1}$.  Since
$\gamma^{-1}$ is small compared to the duration of the growth phase, the
instability has enough time to develop.  Therefore, we point out that
the field-aligned current-driven instability will start earlier in the
region over the heater-enhanced conductivity patch in the $E$ region of
the auroral ionosphere in course of the growth phase when the FAC system
is being enhanced.  A positive feedback in the magnetosphere-ionosphere
coupling was included in order for the initial enhancement to grow
further on until the critical value of FAC being achieved.

Another scenario is the excitation of the turbulent Alfv\'en boundary
layer (TABL) in the ionospheric Alfv\'en resonator (IAR) developed by
\citet{Trakhtengerts&Feldstein91}.

The IAR instability increment $\gamma$ at an arbitrary ionization value
of the ionospheric $E$ layer is expressed by
(\citep{Trakhtengerts&Feldstein91})
\begin{equation}\label{eq:gamma}
\gamma = -\frac{{\VA}_0}{L}\frac{\frac{\SigmaP}{3\SigmaA}}
{1+\left[\frac{\SigmaP}{3\SigmaA}\right]^2}
\left(1-\frac{\Vth}{V_0}\right)
\end{equation}
where ${\VA}_0$ is the Alfv\'en velocity at the height of maximum plasma
concentration, $L$ is the height of the upper wall of the IAR,
$\SigmaP$ is the height integrated Pedersen conductivity of the
ionosphere, $\SigmaA$ is the wave conductivity of the magnetosphere,
$V_0$ is the convection velocity, and $\Vth$ is the minimum
threshold velocity of the convection which is neccessary for an
instability to develop.  The main TABL parameters estimated by
\citet{Trakhtengerts&Feldstein91} are the following:  threshold velocity
of magnetospheric convection $\Vth~\sim 10^{4}$~cm$\cdot$~s$^{-1}$;
the basic turbulence scale of the Alfv\'en waves is
$\lambda\Sub{\perp}\sim1.5$~km; the density of the energy flux
concentrated in the Alfv\'en waves and in the energetic electrons is
$F=100$~erg$\cdot$~cm$^{-2}$~s$^{-1}$.

It is believed that Alfv\'en wave generation during HF pump experiments
provides a possibility for the instability excitation in the ionospheric
Alfv\'en resonator (IAR) which is bounded from below by the
pump-modified ionospheric $E$ region, and from above by the region of
sharp increase of Alfv\'en velocities at heights up to one Earth radius.
The TABL consists of small-scale Alfv\'en vortices ($l\sim1$--$3$~km)
trapped inside IARs.  The accumulation of energy in the magnetospheric
tail is accompanied by an increase in the laminar magnetospheric
convection which, under specific geophysical conditions, can turn into a
turbulent state.  This mechanism would come into play through the local
``switch-on'' of the TABL in the selected magnetic flux tube footprinted
at \Tromso, due to the dependence of the TABL excitation conditions on
the ionospheric $E$ layer.

The local ``switch on'' of the turbulent boundary layer may promote the
formation of a local magnetospheric current system.  This would consist
of two field-aligned sheet currents on the northward and southward sides
of the heater-modified auroral $E$ region closed via Pedersen currents
in the ionosphere.  The configuration of this current system is similar
to the second configuration (Type II) of \citeauthor{Bostrom64}'s model
\citep{Bostrom64} with driving forces inside this system.  In addition,
it drives a Hall current.  The formation of a local magnetospheric
current system is confirmed by the behavior and features of the auroral
arc near \Tromso.  From optical data obtained during heater-on periods,
the appearance of the local spiral-like or bulge structures in the
auroral arcs, typical for the region of the field-aligned currents, can
be clearly identified in the vicinity of \Tromso
(Plates~\ref{fig:IMAGE_Feb17} and~\ref{fig:IMAGE_Feb16}).

\section{Summary}
\label{sec:summary}

Experimental results from multi-instrument observations during \Tromso
HF pumping experiments in the night-side auroral $E$ region indicate
that the localization and timing of the auroral activations were related
to the injection of powerful HF radio waves transmitted from the \Tromso
heating facility.  The modification of the ionosphere-magnetosphere
coupling leading to the local intensification of the auroral activity
may be attributed to the following facts:
\begin{enumerate}

\item
The transmitted $O$-mode ``heater'' waves reflected from the auroral $E$
region may give rise to an increase of the ionospheric conductivity in
the HF pump-modified $E$ region.  The field-aligned current driven
instability will start earlier in the region over the heater-enhanced
conductivity patch during the growth phase of the auroral substorm when
the FAC's system is being enhanced.  A positive feedback in the
magnetosphere-ionosphere coupling was included in order for the initial
enhancement to grow further on until the critical value of the FAC being
achieved.

\item
Excitation of a turbulent Alfv\'en boundary layer (TABL) can take place
under specific geophysical conditions.  The TABL consists of small-scale
($l\sim1$--$3$~km) Alfv\'en vortices trapped inside an ionospheric
Alfv\'en resonator which is bounded from the bottom by the HF pump
modified $E$ region and from the top by the region of sharp increase in
the Alfv\'en velocity at altitudes up to one Earth radius.  The local
``switch on'' of the TABL in the selected magnetic tube can turn the
laminar magnetospheric convection into a turbulent state.

\item
The local ``switch on'' of the TABL may promote the formation of a local
magnetospheric current system.  It consists of two field-aligned sheet
currents on the northward and southward sides of the heater-modified
auroral $E$ region, closed via Pedersen currents in the ionosphere.

\item
The triggering of local auroral activations by HF pump waves requires
specific geophysical conditions, \viz., the presence of the accumulation
of energy in the magnetospheric tail.  The accumulation of energy is
accompanied by an increase of the laminar convection which manifests
itself in the growth of the electric field, formation of the auroral
electrojet, \etc.  In this manner the energy source for the auroral
activations remains the interaction between the solar wind and the
magnetosphere.
\end{enumerate}

We conclude that the experimental results presented here prove the
active r\^ole of the ionosphere in a substorm process and provide
intriguing evidence that the injection of a powerful HF radio beam into
an auroral sporadic $E$ layer may cause the triggering of local auroral
activations.  Therefore, further experiments of the same character are
called for.


\acknowledgements\nopagebreak
We would like to thank the Director and Staff of the EISCAT Scientific
Association for providing the radar data. EISCAT is an International
Association supported by Finland (SA), France (CNRS), the Federal
Republic of Germany (MPG), Japan (NIPR), Norway (NFR), Sweden (NFR), and
the United Kingdom (PPARC).  The work of the first three Russian authors
(N.~B., V.~K., and T.~B.)  was supported by grants from the Swedish
Institute (SI) within the Visby Programme.  They are also grateful to
the Russian Foundation of Fundamental Research, grant 97-05-65443.  The
Swedish author gratefully acknowledges financial support from the
Swedish Natural Sciences Research Council (NFR).  The first two
(N.B. and V.K.) and sixth author (M.R.) were partly supported by
NATO linkage grant, CLG 975069.
The IMAGE magnetometer
data used in this paper were collected as a
German-Finnish-Norwegian-Polish-Russian-Swedish project conducted by the
Finnish Meteorological Institute.  The first three Russian authors wish
to thank the Swedish Institute of Space Physics (Uppsala Division) for
the assistance and hospitality they enjoyed during their visit there.

The authors thank Rolf Bostr\"om, Thomas Leyser, and Tobia Carozzi for
useful discussions and helpful comments.

\bibliographystyle{agufull}
\bibliography{aari}

\end{document}